\newcommand{\vecj}{\mbox{\boldmath$j$}}
\newcommand{\veck}{\mbox{\boldmath$k$}}
\newcommand{\vecm}{\mbox{\boldmath$m$}}

\newcommand{\vecr}{\mbox{\boldmath$r$}}

\newcommand{\vecv}{\mbox{\boldmath$v$}}

\newcommand{\vecA}{\mbox{\boldmath$A$}}
\newcommand{\vecB}{\mbox{\boldmath$B$}}

\newcommand{\vecD}{\mbox{\boldmath$D$}}
\newcommand{\vecE}{\mbox{\boldmath$E$}}

\newcommand{\vecH}{\mbox{\boldmath$H$}}

\newcommand{\vecnl}{\mbox{\boldmath$0$}}

\newcommand{\vecomega}{\mbox{\boldmath$\omega$}}

\newcommand{\dfd}{{\rm d}}

\newcommand{\half}{\frac{1}{2}}

\newcommand{\ie}{{\em i.e.}}
\newcommand{\eg}{{\em e.g.}}
\newcommand{\etal}{{\em et al.}}

\documentclass[12pt]{article}
\usepackage{graphicx}
\begin{document}

\title{Minimum magnetic energy theorem predicts Meissner effect
in perfect conductors}

\author{Miguel C. N. Fiolhais \\  LIP-Coimbra, Department of
Physics, \\
 University of Coimbra, 3004-516 Coimbra, Portugal \and Hanno Ess\'en\thanks{Corresponding author, e-mail: hanno@mech.kth.se} \\
  Department of Mechanics, KTH, 100 44 Stockholm, Sweden \and  C. Provid\^encia \\
  Centro de F{\'i}sica Computacional, Department of
Physics,\\ University of Coimbra, 3004-516 Coimbra, Portugal }

\maketitle

\begin{abstract}
A theorem on the magnetic energy minimum in a perfect, or ideal,
conductor is proved. Contrary to conventional wisdom the theorem
provides a classical explanation of the expulsion of a magnetic
field from the interior of a conductor that loses its resistivity.
It is analogous to Thomson's theorem which states that static
charge distributions in conductors are surface charge densities at
constant potential since these have minimum energy. This theorem
is proved here using a variational principle. Then an analogous
result for the magnetic energy of current distributions is proved:
magnetic energy is minimized when the current distribution is a
surface current density with zero interior magnetic field. The
result agrees with currents in superconductors being confined near
the surface and indicates that the distinction between
superconductors and hypothetical perfect conductors is artificial.
\end{abstract}

\section{Introduction}
In the literature on superconductivity a distinction between
superconductors and hypothetical perfect conductors is often made.
Even though both have zero resistivity, when penetrated by a
constant external magnetic field, the superconductor expels the
field (the experimentally observed Meissner effect) but, in the
corresponding thought experiment, the perfect conductor does not.
Here we will prove a theorem on the magnetic energy minimum from
which one concludes that energy minimization and expulsion of an
external field is the same thing. Therefore, perfect conductors
would in fact behave like superconductors in their approach to
thermodynamic equilibrium. Of course, perfect conductors that are
not superconductors exist only as thought experiments in
textbooks. Nevertheless, textbook thought experiments should be
correct, and unfortunately this is not the case at the moment.

For simplicity we concentrate here on the basic case of type I
superconductors in a field below the critical field. The outline
of this article is as follows. We first discuss the historical
background that led Meissner and subsequent textbook authors
astray. We then present our derivation of the minimum magnetic
energy theorem from a variational principle, after some
preliminary background on Thomson's theorem and its derivation.
Previous work on the problem is discussed and then an illuminating
example is presented in which the energy reduction can be
calculated explicitly. After that we briefly discuss the physical
mechanism of field expulsion. The conclusions are followed by an
Appendix which gives further motivation and explicit solutions for
simple systems.

\section{Meissner effect, diamagnetism, and field expulsion}
In the original 1933 work \cite{meissner} by Meissner and
Ochsenfeld \footnote{For a translation of their article into
English, see \cite{forrest}.}, they stated, without proof,
argument, or reference, that it was understandable that a magnetic
field would not enter a superconductor, but that the expulsion of
a penetrating field at the phase transition to superconductivity
was unexpected. Since then it has been stated innumerable times,
in textbooks \cite{BKtinkham,BKpoole&al,BKgoodstein,BKschmidtvv}
and pedagogical articles \cite{forrest,brandt}, that the expulsion
of the magnetic field cannot be understood in terms of zero
resistivity alone: superconductors are not just perfect
conductors. The purpose of this article is to show that this is
not correct so we must first introduce the confused and intricate
historical context in which the myth arose.

\subsection{The Bohr - van Leeuwen theorem}
In 1911 Niels Bohr \cite{bohr}, and then in 1921, more thoroughly,
van Leeuwen \cite{van_leeuwen}, studied magnetism from the point
of view of classical statistical mechanics. They reached the
conclusion that dia- and paramagnetism could not be explained by
classical theory. This, so called, Bohr-van Leeuwen theorem is
often discussed in books on the theory of magnetism, in particular
by Van Vleck \cite{BKvanvleck}, but also in more recent texts
\cite{BKaharoni,BKmohn}. The model assumed by these authors is a
system of classical charged particles, with Coulomb interactions,
in a constant external magnetic field. The key observation is that
the external magnetic field does not do work on charged particles
and therefore does not change the energy. Hence no statistical
mechanical response. This may seem watertight but there are
weaknesses. For example, the statistical equilibrium distribution
is not always determined simply by the energy. If there are other
conserved phase space quantities, such as angular momentum, they
may also influence the distribution \cite{BKlandau5}. Dubrovskii
\cite{dubrovskii} has criticized the Bohr-van Leeuwen theorem on
these grounds. The main problem though is the assumption of a
purely \emph{external} field, and we return to this below.
Nevertheless, the theorem is still discussed in the literature
\cite{berger,kumar&kumar,kaplan&mahanti} from time to time.

\subsection{The Larmor precession}
There is another, more fundamental, theorem that is in conflict
with the Bohr-van Leeuwen theorem. Basic Lagrangian dynamics can
be used to show that a system of particles, with the same charge
to mass ratio $q/m$, moving in an axially symmetric external
potential, will rotate (precess) with the Larmor frequency $\Omega
= qB/2mc$, when placed in a weak constant external magnetic field
parallel to the axis. This is Larmor's theorem
\cite{BKlarmor,BKlandau2}. It is easy to show that this precession
causes a circulating current which produces a magnetic field that
screens the external field \cite{essen89}. This in accordance with
Lenz law and is the basis of Langevin's classical derivation of
diamagnetism. Feynman \etal\ \cite{BKfeynman&al2} are aware of this
discrepancy and indicate that the Bohr-van Leeuwen result should
not be applied to systems that are free to rotate.

\subsection{The Fock-Darwin Hamiltonian}
The story is, however, even more complicated. A lot of thought has
gone into understanding the physics behind the Bohr-and Leeuwen
theorem. After all, the external field makes free charged
particles circulate with the cyclotron frequency $\omega=qB/mc$,
twice the Larmor frequency. Towards the edges of the system this
should give rise to a net current. It was then realized that the
forces (or the wall) confining the system has the effect of
producing a counter current around the edge that cancels this edge
effect. The exact canceling of the diamagnetic effect by the edge
current, however, contradicts Larmor's theorem, so the situation
is a bit confusing. More detailed studies of these effects have
been made using the model system of independent charged particles
confined axially by a parabolic potential in a constant external
field parallel to the axis. The Hamiltonian for this model is
called the Fock-Darwin Hamiltonian \cite{fock,darwin3}. In recent
years it has been applied to quantum dots.

One notes that the complete diamagnetism of superconductors is in
conflict with the Bohr-van Leeuwen theorem, but not with Larmor's
theorem. In 1938 Welker \cite{welker} showed, quantum
mechanically, that in a free electron gas, confined in a cylinder,
the diamagnetic contribution from Larmor rotation currents is
precisely canceled by the paramagnetic effect on the magnetic
moments due to orbital angular momenta. This fact led Welker to
predict an energy gap in superconductors since this would freeze
the paramagnetic degree of freedom. In spite of all these problems
with understanding the complete diamagnetism this aspect did not
worry Meissner. After all, all metals initially exclude an applied
magnetic field by appropriate induced eddy currents. Due to
resistance these subside, but in ideal conductors they won't. Only
the expulsion of an already present magnetic field at the phase
transition was therefore considered unexpected by Meissner and
Ochsenfeld.

\subsection{The Darwin Hamiltonian}
We note that the findings from the Bohr-van Leeuwen theorem, the
Fock-Darwin Hamiltonian, and even Larmor's theorem, are not
relevant to large systems of charged particles. The reason is that
considering only an external magnetic field is wrong. Unless the
charged particles are all at rest they will be sources of magnetic
fields as they move. When there are many particles this effect,
which necessarily is small in few-body systems, easily becomes
very large \cite{essen96,essen97,essen99,essen06}. For realistic
results one must therefore include the internal magnetic effects
in the relevant dynamics. This was first done by Darwin
\cite{darwin}. The Darwin Lagrangian and corresponding Hamiltonian
can be used to obtain good quantitative understanding of the
complete diamagnetism, as well as the related effect of rotation
called the London moment \cite{essen05}.

\subsection{Thermodynamics and expulsion}
Already in 1934 Gorter and Casimir \cite{gorter&casimir} made  a
thorough analysis of superconductors from the point of view of
classical thermodynamics. One result of these studies is that the
flux expulsion is necessary if the superconducting state is to be
a thermal equilibrium state. If the final situation depends on
which order the phase transition and the external magnetic field
appeared one would have a case of hysteresis, the system would
remember its past. From this point of view flux expulsion is a
natural consequence of approach to thermal equilibrium.

This, of course, does not in itself reveal the actual physical
mechanism of the expulsion. This is not a subject that has been
much discussed in the literature with the notable exception of
Hirsch. In a number of papers Hirsch (see \eg\
\cite{hirsch0,hirsch1,hirsch2,hirsch3}) speculates on the physical
mechanism, and points out that the expulsion is not explained by
BCS theory. Forrest \cite{forrest} argues that conservation of
flux through an ideal current loop, as well as plasma physics
results on the freezing-in of magnetic field lines, prove that
classical electromagnetism cannot explain the expulsion. This will
be refuted below.

\section{Energy minimum theorems}
Electromagnetic energy can be written in a number of different
ways. Here we will assume that there are no microscopic dipoles so
that distinguishing between the $\vecD, \vecH$ and $\vecE, \vecB$
fields is unnecessary. That this is valid when treating the
Meissner  effect in type I superconductors is stressed by Carr
\cite{carr}. Relevant energy expressions are then,
\begin{eqnarray}\label{eq.em.energy.field.form}
E_{\rm e} + E_{\rm m} &=& \frac{1}{8\pi}\!\int (\vecE^2 + \vecB^2)\dfd V \\
\label{eq.em.energy.source.pot.form} &=& \half\int
\left(\varrho\phi +
\frac{1}{c}\vecj\cdot\vecA\right)\dfd V  = \\
\label{eq.em.energy.source.only.form}   &=& \half\int\!\!\int
\left(\frac{\varrho(\vecr)\varrho(\vecr') +
\frac{1}{c^2}\vecj(\vecr)\cdot\vecj(\vecr')}{|\vecr-\vecr'|}
\right)\dfd V\dfd V' .
\end{eqnarray}
Here the first form is generally valid while the two following
assume quasi statics, \ie\ essentially negligible radiation.

According to Thomson's theorem, given a certain number of
conductors, each one with a given charge, the charges distribute
themselves on the conductor surfaces in order to minimize the
electrostatic energy \cite{BKjackson3}. Even though J.\ J.\
Thomson did not present a formal mathematical proof for his
result, the proof of the theorem may be found with great detail in
\cite{BKcoulson,BKpanofsky,BKlandau8} and also on its differential
form in \cite{bakhoum}. A different approach not found in the
literature, based on the variational principle, is presented here.
Thomson's result is widely known and has turned to be very useful
when applied in several distinct situations. For instance, it was
used to determine the induced surface density \cite{donolato}, and
in the tracing and the visualization of curvilinear squares field
maps \cite{bakhoum}. Other applications range from interesting
teaching tools \cite{brito} to useful computational methods such
as Monte Carlo energy minimization \cite{sancho}.

For resistive media currents and magnetic fields always dissipate
but for perfect conductors we will here prove an analogous theorem
for the magnetic field: \emph{The magnetic field energy is
minimized by surface current distributions such that the magnetic
field is zero inside the perfectly conducting bodies, and the
vector potential is parallel to their surfaces.} Since energy
conservation is assumed, it is clearly only valid for ideal
conductors; otherwise steady current exist only if there is an
electric field. Previously similar result have appeared in the
literature \cite{karlsson,badiamajos} and we discuss those below.

\subsection{Thomson's theorem}
In equilibrium, the electrostatic energy functional for a system
of conductors surrounded by vacuum, may be written as\footnote{The
ultimate motivation for this specific form of the energy
functional, and the corresponding one in the magnetic case, is
that they lead to the most elegant final equations.},
\begin{equation}\label{eq.elstat.energy.miguel.form}
E_{\rm e} = \int_{V} \left[  \varrho \phi - {1 \over 8\pi} \left
(\nabla \phi \right ) ^2 \right ] \dfd V ,
\end{equation}
by combining the electric parts of (\ref{eq.em.energy.field.form})
and (\ref{eq.em.energy.source.pot.form})  and using
$\vecE=-\nabla\phi$. We now split the integration region into the
volume of the conductors, $V_{\rm{in}}$, the exterior volume,
$V_{\rm{out}}$, and the boundary surfaces $S$,
\begin{equation}\label{eq.Ee.split.vol}
E_{\rm e}  =  \int_{V_{\rm{in}}} \left[ \varrho \phi - {1 \over
8\pi} \left (\nabla \phi \right )^2 \right ] \dfd V
-\int_{V_{\rm{out}}} {1 \over 8\pi} \left (\nabla \phi \right )^2
\dfd V  + \int_{S} \sigma \phi \,\dfd S ,
\end{equation}
where $\sigma$ is the surface charge distribution.

We now use, $(\nabla \phi )^2 = \nabla \cdot (\phi \nabla \phi)-
\phi \nabla^2 \phi$, and rewrite  the divergencies using Gauss
theorem. The energy functional then becomes,
\begin{eqnarray} E_{\rm e} = \int_{V_{\rm{in}}} \left [ \varrho \phi +
{1 \over 8\pi}  \phi\, \nabla^2 \phi  \right ] \dfd V
    -\int_{V_{\rm{out}}} {1 \over 8\pi} \phi\, \nabla^2 \phi \,
\dfd V \nonumber\\
+\int_{S} \left [ \sigma \phi - {1 \over 8\pi} \phi\, \hat{n}
\cdot \left( \nabla^+ \phi - \nabla^- \phi \right )\right ] \,\dfd
S ,
\end{eqnarray}
where $\nabla^+$ and $\nabla^-$ are the gradient operators at the
surface  in the outer and inner limits, respectively. Since the
total charge in the conductor is limited and constrained to the
conductor volume and surface, one must include this restriction by
introducing a Lagrange multiplier $\lambda$. Therefore the
infinitesimal variation of the energy is,
\begin{eqnarray}
\delta E_{\rm e} &=&  \int_{V_{\rm{in}}} \left [ \delta \phi
\left( \varrho + \frac{1}{4\pi} \nabla^2 \phi \right) + \delta
\varrho \left ( \phi - \lambda \right ) \right ] \dfd V
    +\int_{V_{\rm{out}}} \delta \phi\, \frac{1}{4\pi}
\nabla^2 \phi\; \dfd V \nonumber \\
&&+ \int_{S} \left \{ \delta \phi \left[ \sigma + \frac{1}{4\pi}
\hat{n} \cdot \left( \nabla^+ \phi - \nabla^- \phi \right )
\right] + \delta \sigma \left( \phi - \lambda \right) \right \}
\,\dfd S.
\end{eqnarray}
From this energy minimization, the Euler-Lagrange equations
become:
\begin{equation}
V_{\rm{in}} \left\{
\begin{array}{l}
\nabla^2 \phi = -4\pi \varrho \\
\phi = \lambda \\
\end{array} \right.
\label{potencial}
\end{equation}
\begin{equation}
S \,\,\,\,\left\{
\begin{array}{l}
\sigma = - \frac{1}{4\pi} \hat{n} \cdot
\left( \nabla^+ \phi - \nabla^- \phi \right ) \\
\phi = \lambda \\
\end{array} \right.
\label{surface}
\end{equation}
\begin{equation}
V_{\rm{out}}\hskip 0.5cm \nabla^2 \phi = 0 \label{laplace}
\end{equation}
According to these equations the potential is constant inside the
conductor in the minimum energy state. This means the electric
charge distribution is on the surface and in an equipotential
configuration. This concludes the proof of Thomson's theorem.

\subsection{Minimum magnetic energy theorem}\label{sec.min.mag.energy}
A similar procedure will now be applied to the magnetic field. We
write the magnetic energy functional for a time independent
magnetic field is written as,
\begin{equation}\label{eq.mag.energy.functional}
E_{\rm m} =  \int_{V} \left [\frac{1}{c} \vecj \cdot \vecA - {1
\over 8\pi} \left (\nabla \times \vecA \right ) ^2  \right ]\dfd V
,
\end{equation}
\ie\ as two times the form (\ref{eq.em.energy.source.pot.form}) of
the magnetic energy minus the form
(\ref{eq.em.energy.field.form}), using $\vecB=\nabla\times\vecA$.
As before we split the volume into the volume interior to
conductors, the exterior vacuum, and the surface at the
interfaces, and write,
\begin{eqnarray} \nonumber
E_{\rm m}  &=& \int_{V_{\rm{in}}} \left [ \frac{1}{c} \vecj \cdot
\vecA - {1 \over 8\pi} \left (\nabla \times \vecA \right ) ^2
\right ] \dfd V \\ & & - \int_{V_{\rm{out}}} {1 \over 8\pi} \left
(\nabla \times \vecA \right ) ^2 \dfd V +\frac{1}{c} \int_{S}
\veck \cdot \vecA \,\dfd S ,
\end{eqnarray}
where $\veck$ is the surface current density. We now use the
identity,
\begin{equation}\label{eq.vector.identity.A}
\left (\nabla \times \vecA \right ) ^2 = \nabla \cdot \left [\vecA
\times \left(\nabla \times \vecA \right )\right ] +\vecA \cdot
\left [\nabla \times \left(\nabla \times \vecA \right ) \right ] ,
\end{equation}
and then use Gauss theorem to rewrite the divergence terms. The
energy functional then becomes:
\begin{eqnarray}
E_{\rm m} &=&  \int_{V_{\rm{in}}} \left \{ \frac{1}{c} \vecj \cdot
\vecA - {1 \over 8\pi}  \vecA \cdot \left [\nabla \times
\left(\nabla \times \vecA \right ) \right ]  \right \} \dfd V
\nonumber \\ \label{variational} & & -\int_{V_{\rm{out}}} {1 \over
8\pi}  \vecA \cdot \left [\nabla
\times \left(\nabla \times \vecA \right ) \right ] \dfd V  \\
\nonumber &&+ \int_{S} \left \{ \frac{1}{c} \veck \cdot \vecA - {1
\over 8\pi} \vecA \cdot \left[ \hat{n} \times \left( \nabla^+
\times \vecA - \nabla^- \times \vecA \right) \right] \right \}
\,\dfd S .
\end{eqnarray}
As in the electric field case, some constraints should be imposed.
In particular, due to charge conservation, the electric current
must  obey the continuity equation both in volume and at the
surface \cite{mcallister}:
\begin{eqnarray}
 \nabla \cdot \vecj &=& 0  \\
 \nabla_S \cdot \veck &=& 0
\end{eqnarray}
where $\nabla_S$ is the surface gradient operator. Notice that the
constraints are local, not global. In other words, the Lagrange
multiplier is not constant but a space function $\lambda(\vecr)$.
Therefore, infinitesimal variation of the magnetic energy gives,
\begin{eqnarray} \nonumber
&& c\, \delta E_{\rm m} =  \int_{V_{\rm{in}}} \left \{ \delta
\vecA \cdot \left(  \vecj - \frac{c}{4\pi}\left [\nabla \times
\left(\nabla \times \vecA \right ) \right ] \right) + \delta \vecj
\cdot \left( \vecA-\nabla\lambda \right) \right \} \dfd V  \\
 &&-\int_{V_{\rm{out}}} \delta \vecA \cdot
\frac{c}{4\pi}\left [\nabla \times \left(\nabla \times \vecA
\right ) \right ] \dfd V  \\&& + \int_{S} \left\{ \delta \vecA
\cdot \left[  \veck + \frac{c}{4\pi} \hat{n} \times \left(
\nabla^+ \times \vecA - \nabla^- \times \vecA \right ) \right] +
 \delta \veck \cdot \left( \vecA - \nabla_S \lambda
\right) \right \} \,\dfd S . \nonumber
\end{eqnarray}
Equating this to zero we find that,
\begin{equation}
V_{\rm{in}}\,\, \left\{
\begin{array}{l}
\nabla \times \left(\nabla\times \vecA\right)=\nabla \times
\vecB= \frac{4\pi}{c} \vecj \; , \\
\vecA = \nabla\lambda  \; , \\
\end{array} \right.
\label{potencial2}
\end{equation}
and,
\begin{equation}
S \,\,\,\,\left\{
\begin{array}{l}
\veck =   \frac{c}{4\pi} \hat{n} \times
\left( \nabla^+ \times \vecA - \nabla^- \times \vecA \right ) \; , \\
\vecA = \nabla_S\lambda \; ,\\
\end{array} \right.
\label{surface2}
\end{equation}
and,
\begin{equation}
V_{\rm{out}}\;\;\;  \nabla \times \left(\nabla\times
\vecA\right)=\nabla \times \vecB = 0 , \label{laplace_mag}
\end{equation}
are the Euler-Lagrange equations for this energy functional.

According to eq.\ (\ref{potencial2})
$\vecB=\nabla\times\nabla\lambda=\vecnl$, so the magnetic field
must be zero in $V_{\rm{in}}$. Consequently also the volume
current density is zero, $\vecj=\vecnl$, inside the conductor, in
the minimum energy state.

\subsubsection{Surface currents} Now consider the results for the
surface, eq.\ (\ref{surface2}). Our results from $V_{\rm{in}}$
show that $\nabla^- \times \vecA=\vecnl$, so the equation reads,
\begin{equation}\label{eq.on.surface}
\veck =   \frac{c}{4\pi} \hat{n} \times \left( \nabla^+ \times
\nabla_S\lambda  \right ).
\end{equation}
Let us introduce a local Cartesian coordinate system with origin
on the surface, such that the surface is spanned by $\hat{x},
\hat{y}$ with unit normal $\hat{n}=\hat{z}=\hat{x}\times\hat{y}$.
Assuming that the surface is approximately flat we then have that,
\begin{equation}\label{eq.grad.operators}
\nabla_S = \hat{x}\frac{\partial}{\partial x} +
\hat{y}\frac{\partial}{\partial y},\;\;\mbox{and,}\;\; \nabla^+
=\hat{x}\frac{\partial}{\partial x} +
\hat{y}\frac{\partial}{\partial y} +
\hat{n}\frac{\partial}{\partial z^+} =\nabla_S
+\hat{n}\frac{\partial}{\partial z^+}.
\end{equation}
Since, $\vecA = \nabla_S \lambda$, the vector potential is tangent
to the conducting surface and we get,
\begin{eqnarray}
& \displaystyle \frac{4\pi}{c}\veck = \hat{n}\times \left[
\left(\nabla_S +\hat{n}\frac{\partial}{\partial z^+} \right)
\times \nabla_S \lambda(x,y,z)  \right] &  \\ & \displaystyle =
\hat{n}\times \left[  \hat{n}\frac{\partial}{\partial z^+} \times
\left( \hat{x}\frac{\partial\lambda}{\partial x} +
\hat{y}\frac{\partial\lambda}{\partial y} \right)  \right]  =
\hat{n}\times \left(  \hat{n}\frac{\partial}{\partial z^+}
 \times \vecA  \right), &
\end{eqnarray}
for the surface current density. Rewriting the triple vector
product we find,
\begin{equation}\label{eq.triple.prod.k}
\frac{4\pi}{c}\veck = \frac{\partial}{\partial z^+}
\left[(\hat{n}\cdot\vecA)\hat{n} -
(\hat{n}\cdot\hat{n})\vecA\right] = -\frac{\partial\vecA}{\partial
z^+}.
\end{equation}
so the surface current density is parallel to the outside normal
derivative of the vector potential. We note that this agree with
the well known result \cite{BKlandau8},
\begin{equation}\label{eq.surf.curr.mag.field}
\frac{4\pi}{c}\veck = \hat{n} \times (\vecB_+ - \vecB_-)
\end{equation}
for the case of zero interior field ($\vecB_- =\vecnl$).

\subsubsection{External fields} Even though the proof of the
theorem does not include the presence of an external magnetic
field, the result can be generalized. Assume that the magnetic
field is produced by currents at infinity. These currents demand a
non-zero surface integral at infinity in eq.\ (\ref{variational})
which does not affect the final result, however. Alternatively one
can include large perfectly conducting Helmholtz coils (tori) in
the system and let the rest of the system be small and situated in
the constant field region of the coils. In section
\ref{sec.sphere.perf.cond.in.ext.field} we instead treat this case
by means of an illuminating example.

\subsection{Previous work}
The fact that there is an energy minimum theorem for the magnetic
energy of ideal, or perfect, conductors, analogous to Thomson's
theorem, is not entirely new. In an interesting, but difficult and
ignored, article by Karlsson \cite{karlsson} such a theorem is
stated. Karlsson, however, restricts his theorem to conductors
with holes in them. In the electrostatic case charge conservation
prevents the energy minimum from being the trivial zero field
solution. In the magnetic ideal conductor case a corresponding
conservation law is the conservation of magnetic flux through a
hole \cite{dolecek&delaunay}. But, as long as {\it one} conductor of the system has a hole
with conserved flux there will be a non-trivial magnetic field. To
require that all conductors of the system have holes, as Karlsson
does, seems to us unnecessarily restrictive. One of his main
results is that a the current distribution on a superconducting
torus minimizes the magnetic energy.

A result by Bad{\'i}a-Maj{\'o}s \cite{badiamajos} comes even
closer to our own and we outline it here. The current density is
assumed to be of the form,
\begin{equation}\label{eq.curr.dens.parts}
\vecj = q n \vecv ,
\end{equation}
where $q$ is the charge of the charge carriers and $n$ is their
number density. The time derivative is then given by,
\begin{equation}\label{eq.time.deriv.j}
\frac{\dfd \vecj}{\dfd\, t} =\frac{qn}{m}\, m \frac{\dfd
\vecv}{\dfd\, t} =\frac{qn}{m} \left(q\vecE +
\frac{q}{c}\vecv\times\vecB \right) = \frac{q^2 n}{m} \vecE +
\frac{q}{mc} \vecj\times\vecB ,
\end{equation}
assuming that only the Lorentz force acts (ideal conductor). We
now recall Poynting's theorem \cite{BKlandau2} for the time
derivative of the field energy density of a system of charged
particles,
\begin{equation}\label{eq.poyntings.theorem}
\frac{\dfd }{\dfd\, t} \left(\frac{\vecE^2 + \vecB^2}{8\pi}
\right) = -\vecj\cdot\vecE -\nabla\cdot\left(\frac{c}{4\pi}
\vecE\times\vecB\right).
\end{equation}
The first term on the right hand side normally represents
resistive energy loss. Here we use the result for $\vecE$ from
eq.\ (\ref{eq.time.deriv.j}),
\begin{equation}\label{eq.E.from.j.and.j.dot}
\vecE = \frac{m}{q^2 n} \frac{\dfd \vecj}{\dfd\, t}
-\frac{1}{qnc}\vecj\times\vecB,
\end{equation}
and get,
\begin{equation}\label{eq.E.dot.j}
\vecj\cdot\vecE =\frac{m}{q^2 n}\, \frac{\dfd \vecj}{\dfd\,
t}\cdot\vecj = \frac{\dfd}{\dfd\, t} \left(\frac{m}{2 q^2 n}
\vecj^2 \right).
\end{equation}
This is thus the natural form for this term for perfect
conductors. We insert it into (\ref{eq.poyntings.theorem}),
neglect radiation, and assume that $\vecE^2 \ll \vecB^2$. This
gives us,
\begin{equation}\label{eq.time.deriv.const.quant}
\frac{\dfd}{\dfd\, t} \left(\frac{1}{8\pi}\vecB^2 + \frac{m}{2 q^2
n} \vecj^2 \right) = 0.
\end{equation}
Finally inserting, $\vecj =(c/4\pi) \nabla\times\vecB$, here,
gives,
\begin{equation}\label{eq.energy.for.perf.cond.B}
E_B =\frac{1}{8\pi}\left[ \int_V \vecB^2 \dfd V + \int_{V_{\rm{in}}}
\frac{mc^2}{4\pi q^2 n} (\nabla\times\vecB)^2 \dfd V \right],
\end{equation}
for the conserved energy, after integration over space and time.

Bad{\'i}a-Maj{\'o}s \cite{badiamajos} then notes that this energy
functional implies flux expulsion from superconductors. Variation
of the functional gives the London equation \cite{london&london},
\begin{equation}\label{eq.london.eq}
\vecB + \frac{1}{4\pi}\frac{mc^2}{q^2 n}
\nabla\times(\nabla\times\vecB) = \vecnl.
\end{equation}
Bad{\'i}a-Maj{\'o}s, however, does not point out that this
classical derivation of flux expulsion is in conflict with text
book statements to the effect that no such classical result
exists. Further work by Bad{\'i}a-Maj{\'o}s \etal\
\cite{badiamajos&al}  on variational principles for
electromagnetism in conducting materials should be noted.

\section{Ideally conducting sphere in external field}
We now know that magnetic energy minimum occurs when current flows
only on the surface and the magnetic field is zero inside. Let us
consider a perfectly conducting sphere in a fixed constant
external magnetic field. Here we will calculate the surface
current needed to exclude the magnetic field from the interior and
how much the energy is then reduced. In order to exclude a
constant external field $\vecB$ from its interior the currents on
the sphere must obviously produce an interior magnetic field
$-\vecB$, thereby making the total field zero in the interior. It
is well known that a constant field is produced inside a sphere by
a current distribution represented by the rigid rotation of a
constant surface charge density.
\label{sec.sphere.perf.cond.in.ext.field}

\subsection{Energy of the external field}
For the total magnetic field to have a finite energy we can not
assume that the constant external field extends to infinity.
Instead of using Helmholtz coils we simplify the mathematics and
produce our external field by a spherical shell of current that is
equivalent to a rigidly rotating current distribution on the
surface. This can be done in practice by having as set of rings
representing closely spaced longitudes on a globe with the right
amount of currents maintained in each of them. Such a spherical
shell of rigidly rotating charge produces a magnetic field that is
constant inside the sphere and a pure dipole field outside the
sphere (see Fig.\ \ref{FieldLinesRotSphere}):
\begin{figure}[ht]
\centering
\includegraphics[width=150pt]{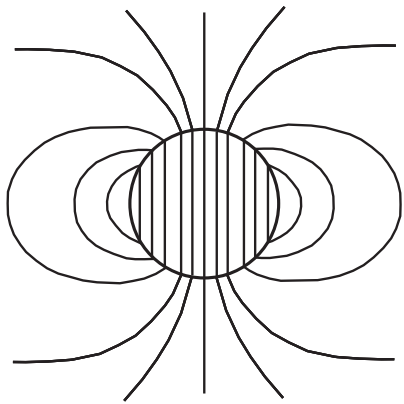}
\caption{\small The field lines of the field of Eq.\
(\ref{eq.mag.field.rot.sphere.shell})  for a nonmagnetic sphere
with a rigidly rotating homogeneous surface charge density.
\label{FieldLinesRotSphere}}
\end{figure}
\begin{equation}\label{eq.mag.field.rot.sphere.shell}
\vecB(\vecr) = \left\{ \begin{array}{ll} \displaystyle \frac{ 2
 \vecm }{R^3}\, & \mbox{for $r\le R$} \\
   &  \mbox{  } \\
\displaystyle \frac{ 3 (\vecm\cdot\hat{r}) \hat{r} - \vecm
}{r^3}\, & \mbox{for $r>R$}
\end{array} \right.
\end{equation}
Here $r=|\vecr|$ and the center of the sphere is at the origin. If
$Q$ is the total rotating surface charge and $\vecomega$ its
angular velocity,
\begin{equation}\label{eq.dipole.vec.of.rot.charge}
\vecm = \frac{Q R^2}{3c} \vecomega ,
\end{equation}
see eq.\ (\ref{eq.middle.field}) below. It is now easy to
calculate the magnetic energy of this field. One finds\footnote{If
(\ref{eq.dipole.vec.of.rot.charge}) is inserted for $\vecm$ here
we get $E_0=(\omega/c)^2 R Q^2/9$ which is equal to $E_m$ of Eq.\
(\ref{eq.energy.thick.rot.shell}) below, for $\xi=1$,
corresponding to surface current only, as it should.},
\begin{equation}\label{eq.energy.mag.field.rot.sphere}
E_0  = \frac{1}{8\pi}\left( \int_{r<R} \vecB^2 \dfd V + \int_{r>R}
\vecB^2 \dfd V
\right)=\left(\frac{2}{3}+\frac{1}{3}\right)\frac{\vecm^2}{R^3} =
\frac{m^2}{R^3}.
\end{equation}
Inside this sphere, which is assumed to maintain a constant
magnetic field in its interior, we now place a smaller perfectly
conducting sphere.

\subsection{Magnetic energy of the two sphere system}\label{sec.two.sphere.syst}
We assume that the small sphere in the middle of the big one has
radius $a<R$ and that it also produces a magnetic field by a
rigidly rotating charged shell on its surface. We denote its
dipole moment by $\vecm_a$ so that its total energy would be,
\begin{equation}\label{eq.energy.inner.alone}
E_a = \frac{\vecm_a^2}{a^3} =  \frac{m_a^2}{a^3},
\end{equation}
far from all other fields, according to our previous result
(\ref{eq.energy.mag.field.rot.sphere}). We now place the small
sphere inside the large one and assume that $\vecm_a$ makes an
angle $\alpha$ with $\vecm$,
\begin{equation}\label{eq.scal.prod.dip.vecs}
\vecm\cdot\vecm_a = m\, m_a \cos\alpha.
\end{equation}
The total energy of the system is now,
\begin{equation}\label{eq.tot.energy.gen.exp}
E =\frac{1}{8\pi}\int (\vecB + \vecB_a)^2 \dfd V = E_0 + E_a + E_i
,
\end{equation}
where the interaction energy is,
\begin{equation}\label{eq.interact.energy.def}
 E_i =\frac{1}{4\pi}\int \vecB \cdot\vecB_a \dfd V
\end{equation}
The integral here must be split into the three radial regions: $0
\le r < a$, $a \le r < R$, and $R \le r$. The calculations are
elementary using spherical coordinates. The contribution from the
inner region is
\begin{equation}\label{eq.interact.energy.1}
 E_{i1} =\frac{1}{4\pi}\int_{r<a} \vecB \cdot\vecB_a \dfd V =
 \frac{4}{3}\frac{mm_a}{R^3} \cos\alpha
\end{equation}
The middle region where there is a dipole field from the small
sphere in the constant field from the big one contributes zero:
$E_{i2}=0$. The outer  region gives $E_{i3} =(2/3) mm_a
\cos\alpha/R^3$. Summing up one finds,
\begin{equation}\label{eq.interact.energy.expl}
E_i = 2 \frac{m m_a}{R^3} \cos\alpha ,
\end{equation}
for the magnetic interaction energy of the two spheres.

\subsection{Minimizing the total magnetic energy}
The total magnetic energy of the system discussed above is thus,
\begin{equation}\label{eq.tot.energy.as.func}
E_0+E_a+E_i = E(m_a,\alpha) = \frac{m^2}{R^3} + \frac{m_a^2}{a^3}
+ 2 \frac{m m_a}{R^3} \cos\alpha .
\end{equation}
We assume that $m$ and $m_a$ are positive quantities. This means
that as a function of $\alpha$ this quantity is guarantied to have
its minimum when $\cos\alpha = -1$, \ie\ for $\alpha = \pi$. Thus,
at minimum, the dipole of the inner sphere has the opposite
direction to that of the constant external field.

Now assuming $\alpha = \pi$ we can look for the minimum as a
function of $m_a$. Elementary algebra shows that this minimum is
attained for,
\begin{equation}\label{eq.min.ma}
m_a =\left( \frac{a}{R} \right)^3 m\, \equiv m_{\rm min}.
\end{equation}
The magnetic field in the interior of the inner sphere is then,
\begin{equation}\label{eq.mag.field.interior.min}
\vecB + \vecB_a = \frac{ 2\vecm }{R^3} + \frac{ 2\vecm_a }{ a^3 }
= \left( \frac{2 m }{ R^3 } - \frac{ 2 m_{\rm min}  }{ a^3 }
\right)
 \frac{ \vecm }{m} = \vecnl ,
\end{equation}
so it has been {\it expelled}. The minimized energy
(\ref{eq.tot.energy.as.func}) of the system is found to be,
\begin{equation}\label{eq.minimum.energy.two.sphere}
E_{\rm min} = E(m_{\rm min},\pi) = \frac{m^2}{R^3} \left[ 1 -
\left( \frac{a}{R} \right)^3 \right].
\end{equation}
The relative energy reduction is thus given by the volume ratio of
the two spheres. Using $\vecB=2\vecm/R^3$ from
(\ref{eq.mag.field.rot.sphere.shell}) the energy lowering can also
be expressed as follows:
\begin{equation}\label{eq.energy.lowering.of.expulsion}
E_0 - E_{\rm min} = \frac{m^2}{R^3}\frac{a^3}{R^3}
=\frac{\vecB^2}{4} a^3 = 3\left( \frac{\vecB^2}{8\pi}\right)
\left( \frac{4\pi a^3}{3} \right).
\end{equation}
This result is independent of the radius of the big sphere
introduced to produce the constant external field. It shows that
the energy lowering corresponds to three times the external
magnetic energy in the volume $4\pi a^3/3$ of the perfectly
conducting sphere.

\section{The mechanism of flux expulsion}
Assume that a resistive metal sphere is penetrated by a constant
magnetic field. Lower the temperature until the resistance
vanishes. How does the metal sphere expel the magnetic field, or
equivalently, how does it produce surface currents that screen the
external field? According to Forrest \cite{forrest} this can not
be understood from the point of view of classical electrodynamics
since in a perfectly conducting medium the field lines must be
frozen-in. This claim is motivated thus: When the resistivity is
zero there can be no electric field according to Ohm's law, since
this law then predicts infinite current. But if the electric field
is zero the Maxwell equation, $\nabla\times\vecE -
\partial\vecB/\partial\, t =\vecnl$, requires that the time
derivative of the magnetic field is zero. Hence it must be
constant.

Ohm's law is, however, hardly applicable in this situation. The
system of charged particles undergoes thermal fluctuations and
these produce electric and magnetic fields. The electric fields
accelerate charges according to the Lorentz force law. In the
normal situation these currents stay microscopic. When there is an
external magnetic field present the overall energy is lowered if
these microscopic currents correlate and grow to exclude the
external field. According to standard statistical mechanics the
system will eventually relax to the energy minimum state
consistent with constraints. We note that Alfv\'en and
F\"althammar \cite{BKalfven} state that "in low density plasmas
the concept of frozen-in lines of force is questionable".

Another argument by Forrest \cite{forrest} is that the magnetic
flux through a perfectly conducting current loop is conserved.
Since one can imagine arbitrary current loops in the metal that
just lost its resistivity with a magnetic field inside, the field
must remain fixed, it seems. Is is indeed correct that the flux
through ideal current loops is conserved, but the actual physical
current loop will not remain intact unless constrained by
non-electromagnetic forces. On a loop of current that encloses a
magnetic flux there will be forces that expand it
\cite{BKlandau8}. This is the well know mechanism behind the rail
gun, see \eg\ Ess\'en \cite{essen09}. All the little current loops
in the metal will thus expand until they come to the surface where
the expansion stops. In this way the interior field is thinned out
and current concentrates near the surface. So, the flux is
conserved through the loops, but the loops expand.

\section{Conclusions}
We have now reached the conclusion that the Meissner effect, both the perfect diamagnetism and the flux expulsion, are consequences of classical electrodynamics, together with reasonable assumptions about the system of charged particles. The reason that this has not been recognized before is a widespread ignorance of the properties and effects of the magnetic interaction energy between moving charged particles. The majority of the physics community tends to think of magnetism as either due to microscopic dipoles in a medium or as due to an imposed external magnetic field. That the translational motion of the charged particles of the system also produces magnetic fields, which in turn change their motion, is normally neglected. When this can not be neglected one must find a self consistent simultaneous solution for currents and magnetic fields \cite{essen04}. More awareness of these facts will hopefully clear up a lot of confusion in the future.

\appendix
\section{Appendix}
\subsection{Magnetic energy minimization in simple one degree of
freedom model systems} We now define simple one degree of freedom
model systems and use them to illustrate how the minimum magnetic
energy theorem works. We take systems in which all energies can
be calculated exactly so that energy minimization amounts to
minimizing a function of a single variable. The systems are both
related to a system used by Brito and Fiolhais \cite{brito} to
study electric energy.

\subsubsection{Magnetic energy of coaxial cable}
The minimum magnetic energy theorem can be illustrated in such a
simple system as a coaxial cable. The cable can be modelled by an
outer cylindrical conducting shell with radius $b$, carrying an
electric current $I$, and a concentric solid cylindrical conductor
with radius $a < b$, carrying the same electric current in the
opposite direction. We now divide the inner current into a surface
current $I_s = (1-\eta)I$ at $\rho = a$ and a bulk current $I_v=\eta
I$ in the inner volume, $0\le\rho<a$, so that $I=I_s+I_v$. We use
cylindrical coordinates, $\rho, \varphi, z$, and obtain the
magnetic field,
\begin{equation}
\vecB = \frac{2 I}{c}\cdot \left\{
\begin{array}{ll} \displaystyle
\frac{ \eta  \rho}{ a^2}\, \hat{\varphi}&\;\; 0 \le \rho <a \\
    & \\ \displaystyle
\frac{1}{\rho}\, \hat{\varphi}&\;\; a \le  \rho \le b \\
 & \\ \displaystyle
0 &\;\; b < \rho \\
\end{array} \right.
\label{surface3}
\end{equation}
using Amp\`ere's law. Thus, the magnetic field energy for a length
$L$ of the cable is given by:
\begin{eqnarray}
E_{\rm m} & = & \frac{1}{8\pi}\int_{V} \vecB^2 \dfd V \nonumber\\
 & = & \frac{1}{8\pi} \left(\frac{2 I}{c}\right)^2  \left[ \int_{0}^a \left ( \frac{ \eta \rho}{ a^2} \right )^2
  L 2 \pi \rho \, \dfd  \rho +
 \int_{a}^b \left ( \frac{ 1}{\rho} \right )^2 L
2 \pi \rho \, \dfd  \rho \right] \nonumber \\
 & = & \frac{L I^2}{c^2}\left[ \frac{\eta^2}{4} + \ln\!\left(\frac{b}{a}\right)
 \right].
\end{eqnarray}
This magnetic energy reaches its minimum for zero bulk current,
$\eta=0$, corresponding to surface current only and zero field for
$0 \le \rho <a$.

\subsubsection{Current in sphere due to rigidly rotating charge}
Consider an ideally conducting sphere of radius $R$. Assume that
there is a circulating current in the sphere which can be seen as
the rigid rotation of a charge $Q$ evenly distributed in the thick
spherical shell between $r=a <R$ and $r=R$. The charge density,
\begin{equation}\label{eq.charge.dens.thich.sphere.shell}
\varrho(\vecr) = \left\{ \begin{array}{ll}\displaystyle 0 & \mbox{for $0 \le r < a$} \\
   &  \mbox{  } \\
\displaystyle \frac{3Q}{4\pi(R^3-a^3)}
& \mbox{for $a\le r \le R$} \\
   &  \mbox{  } \\
   0 & \mbox{for $R < r$}
\end{array} \right.
\end{equation}
is assumed to rotate with angular velocity $\vecomega
=\omega\,\hat{z}$ relative to a an identical charge density of
opposite sign at rest. The current density is then,
\begin{equation}\label{eq.curr.dens.rot}
\vecj(\vecr)=\varrho(\vecr)\,\vecomega\times\vecr,
\end{equation}
and the current, $ I = \frac{Q}{2\pi} \omega , $ passes through a
half plane with the $z$- axis as edge.

The vector potential produced by this current density can be found
using the methods of Ess\'en \cite{essen89}, see also
\cite{essen96,essen06,essen05}. If we introduce $\xi = a/R$ we
find,
\begin{equation}\label{eq.vec.pot.rot.charge.dens.thich.sphere.shell}
\vecA(\vecr) = \frac{Q}{c} (\vecomega\times\vecr) \cdot
\left\{ \begin{array}{ll}\displaystyle  \frac{(1-\xi^2)}{2(1-\xi^3)R} & \mbox{for $0 \le r < a$} \\
   &  \mbox{  } \\
\displaystyle \frac{R^2}{10 (1-\xi^3)} \left(\frac{5}{R^3} -
3\frac{r^2}{R^5} -2\frac{\xi^5}{r^3} \right)
& \mbox{for $a\le r \le R$} \\
   &  \mbox{  } \\  \displaystyle
 \frac{(1-\xi^5)}{5(1-\xi^3)}\frac{R^2}{r^3} & \mbox{for $R < r$}
\end{array} \right.
\end{equation}
The parameter $\xi=a/R$ is zero, $\xi=0$, for a homogeneous ball
of rotating charge, while $\xi=1$ corresponds to a rotating shell
of surface charge. Comparing with the vector potential for a
constant field, $\vecA = \half \vecB_0\times\vecr$ we see that,
\begin{equation}\label{eq.middle.field}
\vecB_0 =  \frac{Q\vecomega}{c R} \frac{(1-\xi^2)}{(1-\xi^3)} =
\frac{Q\vecomega}{c R} \frac{(1+\xi)}{(1+\xi+\xi^2)}
\end{equation}
is the field in the central current free region $0\le r<a$.

\subsubsection{Magnetic energy of rotating spherical shell current}
We now calculate the magnetic energy of this system using the
formula,
\begin{equation}\label{eq.energy.intergral.gen}
E_m = \frac{1}{2c} \int \vecj \cdot \vecA \; \dfd V .
\end{equation}
Performing the integration using spherical coordinates gives,
\begin{equation}\label{eq.energy.thick.rot.shell}
E_m = \left(\frac{R\omega}{c}\right)^2 \frac{Q^2}{R}\; f(\xi) ,
\end{equation}
where,
\begin{equation}\label{eq.f.of.xi}
f(\xi) =
\frac{2+4\xi+6\xi^2+8\xi^3+10\xi^4+5\xi^5}{35(1+\xi+\xi^2)^2},
\end{equation}
is a function of the dimensionless parameter $\xi$.

\begin{figure}[ht]
\centering
\includegraphics[width=230pt]{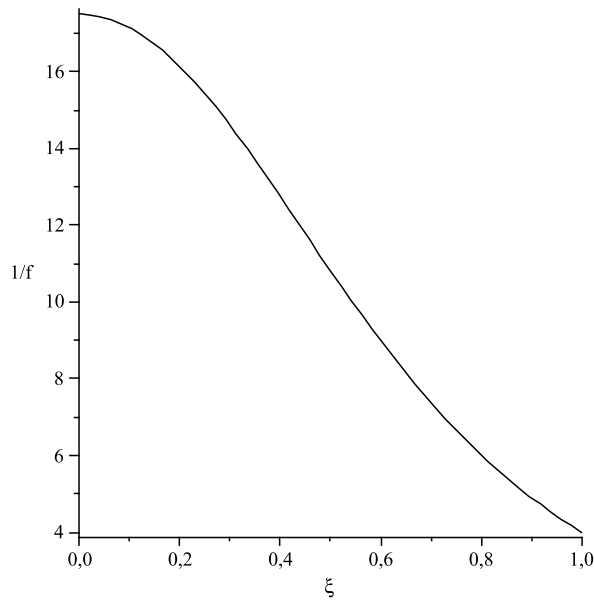}
\caption{\small A graph of the function $1/f(\xi)$ which is
proportional to the Hamiltonian form of the magnetic energy
(\ref{eq.ham.energy.of.model}) of our model system. Note that
$\xi=0$ corresponds to volume (bulk) current and $\xi=1$ to pure
surface current. \label{fofxicurve}}
\end{figure}

This expression for the energy is the Lagrangian form of a kinetic
energy which depends on the generalized velocity
$\omega=\dot\varphi$,
\begin{equation}\label{eq.lagr.energy.thick.rot.shell}
L_m(\dot\varphi) = \frac{R^2}{c^2} \frac{Q^2}{R}\;
f(\xi)\,\dot\varphi^2 .
\end{equation}
Since the generalized coordinate $\varphi$ does not appear in the
Lagrangian $L_m$ the corresponding generalized momentum (the
angular momentum),
\begin{equation}\label{eq.gen.mom}
p_{\varphi}=\frac{\partial L_m}{\partial \dot\varphi} = 2
\frac{R^2}{c^2} \frac{Q^2}{R}\; f(\xi)\,\dot\varphi ,
\end{equation}
is a conserved quantity. The corresponding Hamiltonian form for
the energy is then $H_m = p_{\varphi} \dot\varphi - L_m$,
expressed in terms of $p_{\varphi}$,
\begin{equation}\label{eq.ham.energy.of.model}
H_m(p_{\varphi}) = \frac{c^2}{4} \frac{p_{\varphi}^2}{ Q^2 f(\xi)
R}.
\end{equation}
The function $1/f(\xi)$ is plotted in Fig.\ \ref{fofxicurve}.
We now consider the two energy expressions
(\ref{eq.lagr.energy.thick.rot.shell}) and
(\ref{eq.ham.energy.of.model}) separately.

{\bf Case of constant current:} We first consider the case that
the {\it current,} $I = Q\omega/2\pi$, {\it is constant}. Changing
$\xi$ then means changing the conductor geometry while keeping a
constant total current, or, equivalently, angular velocity
$\omega=\dot\varphi$. One might regard the total current as
flowing in a continuum of circular wires. Changing $\xi$ from zero
to one means changing the distribution of these circular wires
from a bulk distribution in the sphere to a pure surface
distribution, while maintaining constant current. According to a
result by Greiner \cite{BKgreinerCED} a system will tend to {\it
maximize} its magnetic energy when the conductor geometry changes
while currents are kept constant. This has also been discussed in
Ess\'en \cite{essen09}. In conclusion, if currents are kept
constant the magnetic energy
(\ref{eq.lagr.energy.thick.rot.shell}) will tend
(thermodynamically) to a stable equilibrium with at a {\it
maximum} value and we note that this corresponds to a pure surface
current $\xi=1$.

{\bf Case of constant angular momentum:} Assume now that we  pass
to the Hamiltonian (canonical) formalism. Thermodynamically this
type of system should tend to minimize  its phase space energy
(\ref{eq.ham.energy.of.model}) in accordance with ordinary
Maxwell-Boltzmann statistical mechanics. As a function of $\xi$
this Hamiltonian form of the energy clearly has a minimum at $\xi
=1$, see Fig.\ \ref{fofxicurve}, corresponding to pure surface
current. In this case therefore there will be current density only
on the {\it surface} in the energy minimizing state. This is in
accordance with the our minimum magnetic energy theorem. It is
notable that {\it both} the assumption of constant current and the
assumption of constant angular momentum lead to a pure {\it
surface current density} as the stable equilibrium.

\subsection{Explicit solutions with minimum magnetic energy}
Here we present two explicit solutions for the current
distributions and magnetic fields that minimize the magnetic
energy. The solution for a torus has been found several times,
probably first by Fock \cite{fock2}, but, independently, several
times since then, see \eg\
\cite{launay,carter&loh&po,bhadra,haas1,belevitch&boersma,ivaska&al},
so we do not repeat it here. Karlsson \cite{karlsson},  however,
was probably the first to notice that the solution minimizes
magnetic energy for constant flux. Dolecek and de Launay \cite{dolecek&delaunay}
verified experimentally that a type I superconducting torus behaves
exactly as the corresponding classical perfectly diamagnetic
system for field strength below the critical field. Here we treat two simpler cases involving constant
external fields: a cylinder in a transverse magnetic field, and a
sphere.

\subsubsection{Cylinder in external perpendicular magnetic field}
Consider an infinite cylindrical ideal conductor with radius $R$
in a external constant perpendicular magnetic field. To get the
vector potential one must solve the following differential
equation,
\begin{equation}
\nabla \times \vecB = \nabla \times \left ( \nabla \times \vecA \right )= 0 .
\label{curl}
\end{equation}
To solve this one should look for the symmetries of the system. We
assume that the external constant magnetic field points in the
$y$-direction and that the cylinder axis coincides with the
$z$-axis. There will then be no dependence on the $z$-coordinate
so the magnetic field is,
\begin{equation}
\vecB =  \frac{1}{\rho} \frac{\partial
A_z}{\partial \varphi}\,\hat{\rho}
 -  \frac{\partial A_z}{\partial \rho}\,\hat{\varphi} +  \frac{1}{\rho}
  \left ( \frac{\partial  }{\partial \rho} (\rho A_{\varphi}) -
  \frac{\partial A_{\rho}}{\partial \varphi} \right )\hat{z} .
\label{magnetic_2}
\end{equation}
Moreover, due to the symmetry of the system, the $z$-component of
the  magnetic field must be zero,
\begin{equation}
\frac{\partial  }{\partial \rho} (\rho A_{\varphi}) -
\frac{\partial A_{\rho}}{\partial \varphi} = 0 .
\label{equacao}
\end{equation}
These assumptions and constraints transform eq.\ (\ref{curl})
into,
\begin{equation}
\nabla \times \vecB = -  \left ( \frac{1}{\rho^2}
\frac{\partial^2 A_z}{\partial \varphi^2} +  \frac{\partial^2
A_z}{\partial \rho^2} +  \frac{1}{\rho}  \frac{\partial
A_z}{\partial \rho}   \right ) \hat{z}= 0
\end{equation}
which simply is Laplace equation in cylindrical coordinates
($\rho,\varphi,z$).  Before writing down the general  solution,
let's analyze the boundary conditions. As $\rho \rightarrow
\infty$, the magnetic field must approach the external one:
$\vecB_0 = B_0 \hat{y} = B_0 (\hat{\varphi} \cos \varphi
+\hat{\rho} \sin \varphi ) $. Furthermore, since the magnetic
field is zero inside the perfect conductor, one concludes from
eq.\ (\ref{magnetic_2}) that the vector potential vector must be
constant inside the cylinder. Therefore, the solution is,
\begin{equation}
A_z = \textrm{const.} + B_0 \left (  \frac{R^2}{\rho} - \rho \right ) \cos \varphi ,
\end{equation}
for $\rho > R$. The magnetic field
outside the cylinder becomes,
\begin{equation}
\vecB = \hat{\rho}\, B_0 \left ( 1 - \frac{R^2}{\rho^2}
\right)\sin \varphi
 + \hat{\varphi}\, B_0 \left( 1 + \frac{R^2}{\rho^2} \right )\cos \varphi ,
\label{magnetic}
\end{equation}
which implies that,
\begin{equation}
\vecB(\rho=R) = 2 B_0 \cos \varphi \,\, \hat{\varphi}
\label{magnetic8}
\end{equation}
The magnetic field on the cylinder's surface
determines the surface current according to eq.\ (\ref{eq.surf.curr.mag.field}), so we get,
\begin{equation}
\veck =  \frac{c }{2 \pi} B_0  \cos \varphi \,\, \hat{z}.
\label{surfacecurrent}
\end{equation}
The total current obtained through integration of the surface
current is zero as expected, otherwise the energy would diverge. A
more detailed analysis on this problem can be found in
\cite{zhilichev}.

\subsubsection{Superconducting sphere in constant magnetic field}
Similar calculations can be performed for a superconducting sphere
with radius $R$ in a constant external magnetic field pointing in
the direction of the $z$-axis.

Similarly to the cylinder case, eq.\ (\ref{curl}) is simplified a
lot due to the symmetries of the system. Since the
external constant magnetic field points along
the $z$-axis, there won't be any dependencies on the $\varphi$
coordinate and the magnetic field along this coordinate must be
zero. Therefore, the magnetic field simplifies to,
\begin{equation}
\vecB =\frac{1}{r \sin \theta} \frac{\partial} {\partial \theta}
 \left( A_\varphi \sin \theta \right) \hat{r} -  \frac{1}{r}
  \frac{\partial }{\partial r}  \left( A_\varphi r \right) \hat{\theta} ,
\label{magnetic_3}
\end{equation}
where we use spherical coordinates $r, \theta, \varphi$. Again,
using  these assumptions and constraints eq.\ (\ref{curl})
becomes,
\begin{equation}
\nabla \times \vecB = -  \frac{1}{r} \left [
\frac{\partial }{\partial r} \left ( \frac{\partial }{\partial r}
 \left ( rA_\varphi \right ) \right ) + \frac{1}{r} \frac{\partial }{\partial \theta}
 \left ( \frac{1}{\sin \theta} \frac{\partial }{\partial \theta}
 \left ( A_\varphi \sin \theta \right ) \right ) \right ] \hat{\varphi} = 0
\label{curl_2}
\end{equation}
Since the magnetic field is zero inside the sphere, eq.\
(\ref{magnetic_3}) implies that the vector potential has the form,
\begin{equation}
A_\varphi (r < R) = \frac{C}{r \sin \theta} ,
\end{equation}
where $C$ is a constant. To prevent the vector
potential from diverging at $r=0$ and $\theta = 0$, the constant $C$
must be zero. Furthermore, as $r \rightarrow \infty$, the magnetic
field must go to the external field, $\vecB_0 = B_0
\hat{z} = B_0 (\hat{r} \cos \theta -\hat{\theta} \sin \theta ) $.
Therefore, the solution of  eq.\ (\ref{curl_2}) for this case is,
\begin{equation}
A_\varphi (r > R) = \frac{B_0}{2} \left (  r - \frac{R^3}{r^2}
\right ) \sin \theta ,
\end{equation}
which leads to the following magnetic field outside the sphere,
\begin{equation}
\vecB = \hat{r}\, B_0  \left ( 1 - \frac{R^3}{r^3}  \right) \cos
\theta  - \hat{\theta}\, B_0 \left( 1 + \frac{1}{2}
\frac{R^3}{r^3} \right )\sin \theta . \label{magnetic_4}
\end{equation}
The magnetic field at the sphere surface
is thus,
\begin{equation}
\vecB = - \frac{3}{2} B_0  \sin \theta  \,\, \hat{\theta}.
\label{magnetic_5}
\end{equation}
One notes that this is the same field as that of section
\ref{sec.two.sphere.syst} at the surface of the inner sphere when
energy is minimized.

Using eq.\ (\ref{eq.surf.curr.mag.field}), the surface current density becomes,
\begin{equation}
\veck = - \frac{3c}{8\pi} B_0 \sin \theta  \,
\hat{\varphi} \label{current_3}
\end{equation}
Unlike the infinite cylinder in a perpendicular external  field,
the sphere must have a total non-zero electric current, $I =
\frac{3c}{4\pi}R B_0 $, to keep the magnetic field from entering.
A similar approach to this problem can be found in \cite{matute}.
\\ \\
\noindent
{\bf\large Acknowledgements}
H.\ E.\ would like to thank Arne B.\ Nordmark for verifying some of the results of this article by means of finite element computation.



\end{document}